# Predicting User-Interactions on Reddit

Maria Glenski  Tim Weninger
University of Notre Dame
Notre Dame, Indiana 46556
Email: {mglenski, tweninge}@nd.edu

*Abstract*—In order to keep up with the demand of curating the deluge of crowd-sourced content, social media platforms leverage user interaction feedback to make decisions about which content to display, highlight, and hide. User interactions such as likes, votes, clicks, and views are assumed to be a proxy of a content's quality, popularity, or news-worthiness. In this paper we ask: how predictable are the interactions of a user on social media? To answer this question we recorded the clicking, browsing, and voting behavior of 186 Reddit users over a year. We present interesting descriptive statistics about their combined 339,270 interactions, and we find that relatively simple models are able to predict users' individual browse- or vote-interactions with reasonable accuracy.

## I. INTRODUCTION

Social media platforms rely on the collective ratings of its users to shape the manner by which content is be displayed. Users rely on those ratings and rankings to make a wide array of decisions about what to read or watch, which brand to buy, or which hotel to reserve. Most users are passive browsers or "lurkers." Only a handful of users actually contribute ratings, and even fewer still contribute new content [1]–[3]. Because of this asymmetric user behavior, we assume that it is difficult to model the reading, rating, and writing interactions of users on social media platforms.

Social influence biases present an additional challenge to user modelling. When users provide ratings to social posts (or comments), the system's ranking algorithm reorders the content promoting highly rated content over others. Previous studies have found that this so-called *ranking bias effect* encourages herding behavior that results in suboptimal outcomes [4]–[7]. This leads to asymmetric outcomes for posts of identical content [8], [9]. In social *networks*, these algorithmic biases may be responsible for viral information cascades [10], [11] and may play a significant role in the spread of misinformation in social media [12].

The related work in social media analysis relies exclusively on the information provided by the social media platform – typically through an API or by scraping their Web interface. Important questions can certainly be answered in this way, but fine-grained user behavior is not usually made available because of privacy concerns. The absence of a user's precise behavior (*i.e.*, what they read, rated, and wrote) precludes fine-grained user modelling.

In contrast to previous studies, we tracked the online activity of 309 users of the popular social news aggregator Reddit from August 1, 2015 to July 31, 2016. Study participants were recruited through posts to various subreddits, via links from other Web surveys, and locally among the students at the University of Notre Dame. Participants were asked to download a browser extension (available for Firefox and Chrome browsers) that reported usage data of all clicks, pageloads, and votes cast within the reddit.com domain while they were signed into their Reddit accounts. To mitigate the potential for malicious participants, their accounts must have been created at least a week before installation.

Upon installation, the browser extension asked the user to opt-in to the study via an informed consent form; the user could modify their consent or uninstall the extension by clicking a small icon that was embedded next to their username in the top-right of Reddit Web pages. We only recorded activity that occurred while the participant was signed in to their Reddit account and only when their browser was not in private or "incognito" browsing mode. All data was anonymized on the client-side. We did not collect IP addresses, nor did we collect posts or comments made by the user as these could be used to easily de-anonymize the data.

It is important to be mindful of potential sampling bias. The only way to collect this data is for users to self-select, *i.e.*, opt-in, to the project. Self-selection bias, especially in online polling, for example, will often skew responses in favor of the most motivated group. However, we do not expect much bias from self-selection because our system does not ask any questions; it merely observes user behavior. Our recruitment strategy may be affected by undercoverage bias, where certain groups are not included in the sample. So, because of the potential for undercoverage bias, we do not intend for this data to be representative of broad or group-based opinion or popularity. Indeed, the opinions and views expressed on Reddit itself are not broadly representative of the general public.

This dataset contains about 2 million total interactions and may be used to answer interesting questions apart from those addressed in this paper.

We recorded every click and page-load made by participants in the reddit domain that is not publicly linked to their account. So, there are many types of interactions in this dataset. These include: upvote a post, upvote a comment, save a post, or perform any number of other interactions. In the present work we focus specifically on when a user: 1) *upvote*s a post 2) *downvote*s a post 3) *browse*s a post's content, or 4) *browse*s the *comments* section of a post. We say that a user browses a post's content if they click on the post's title (which is always a hyperlink), or if they click to view the content of the post through expander buttons or by other means. Similarly, we say that a user browses the comments section of a post if they, in any way, navigate to the comments section of a post. We examine the activity of our sample of Reddit users with Reddit posts via two complementary tasks:

1) **User Interactions.** First, we examine the effects that ranking bias, score bias, and recency bias have on user behavior when interacting with posts on Reddit. We also extract shallow linguistic features from the post titles to uncover correlations between a post's title and the type of interaction that the post attracts.

2) **Predictability.** Next, we show that the features extracted in the first task can be used to model the activity of individual users. Our relatively straightforward user model is able to predict a user's interaction (or lack thereof) with a new post with surprising accuracy.

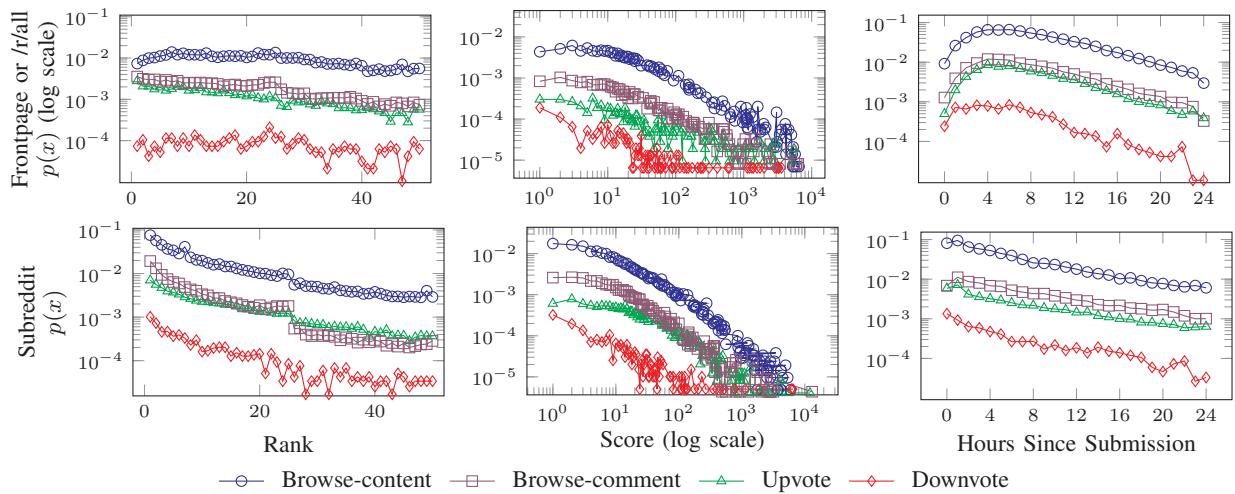

Fig. 1. Probability distributions for each interaction type by rank (left), score (center), and hours since submission (right) at interaction time separated into interactions that occurred from the frontpage or /r/all (top) or a subreddit (bottom) view.

From the original set of 309 respondents, we found that many users did not interact with many posts despite frequently visiting Reddit and (presumably) reading its headlines without ever actually browsing or voting. Because we are unable to track the posts that a user did not browse or vote, we threw out users than had less than 10 total interactions with posts. This resulted in 186 Reddit users with 339,270 browse- or vote-interactions.

The next section discusses a limited number of descriptive statistics that illustrate interesting properties of user behavior. Then, we describe our user model and present the results of experiments that predict future user-interactions. We conclude with the predictions of an example user and discuss the wider implications of our findings.

## II. USER INTERACTIONS

Previous studies have identified the impact that various information curation mechanisms have on the success or failure of submitted content. Collectively we refer to these as *algorithmic design biases*, which include the ranking bias, score bias, and recency effects. These algorithmic design biases are embedded into online media platforms in some way or another as a way to present relevant and popular content. They are especially prevalent on social media platforms like Facebook, Twitter, Youtube, and Reddit, which we highlight in the current work.

Of the 339,270 user interactions we examine, there are 27,618 upvotes, 2,431 downvotes, 265,239 content-views, and 43,982 comment-views. The algorithmic design of Reddit significantly influences the posts that are highly visible and likely candidates for user interaction. Consider, for example, the ranking bias effect, in which higher ranked posts have more visibility and are therefore more likely to be clicked. It is reasonable to assume that Reddit is susceptible to the ranking bias effect, but the severity of its influence is unknown. Figure 1 shows the probability distributions of browse-content, browse-comment, upvote, and downvote interactions. The top row of Fig. 1 shows interactions from the frontpage or /r/all[1], and the bottom row shows only interactions that occurred from within a specific subreddit.

Contrary to our initial hypothesis, we find only a small rank bias effect on the frontpage. The probability of browse interactions, especially, had little drop off as the rank ranged from 1 to 50, even though only 25 posts are listed per page. Within a subreddit view, user interactions did show a more obvious ranking bias effect.

The middle column in Fig. 1 shows the probability distribution (log-log scale) of post scores (*i.e.*, total upvotes-downvotes) at the time of the page load. We do not record a user's upvote as +1 to the score. Note that Reddit frequently hides post-scores of newly submitted content, thus approximately 40% of our recorded interactions had no score present. Of the remaining interactions, we observe an long-tailed distribution wherein posts with the most interactions have a score less than 5. This was expected because the vast majority of posts garner little attention.

---

[1]Reddit's new-default /r/popular frontpage was not present during the data collection period.

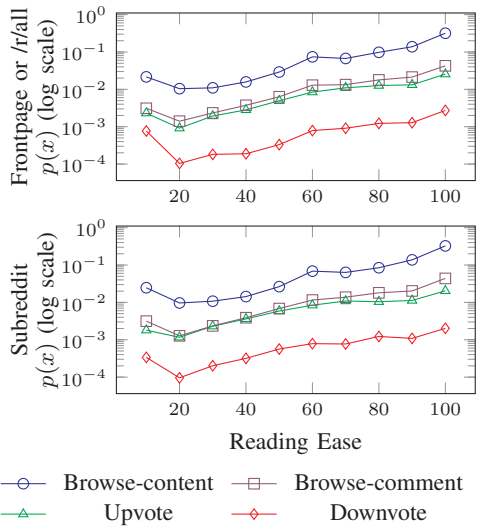

Fig. 2. Probability distributions of Flesch reading ease scores for titles of posts that received content-browses, comments-browses, upvotes, and downvotes where higher reading ease scores depict easier to read titles, separated into interactions that occurred from the frontpage or /r/all (top) or a subreddit (bottom) view.

The rightmost column in Fig. 1 shows the distribution (log scale) of interactions as a function of time (in hours) since the post was submitted. When viewed from the frontpage, the distribution appears to be a right-skewed Gaussian with a median of 4 hours. This seems reasonable because Reddit posts require some time to garner enough votes to appear on a user's frontpage or /r/all. Within a specific subreddit, we see a slightly accelerated linear decay, where most interactions happen within the first hour or two of the post's lifecycle.

User interaction is also guided by the title of a post, which is typically limited to a handful of descriptive words that encourage the user to browse the content, vote, comment or otherwise engage. Using shallow textual features such as the number of syllables, words, sentences, Flesch proposed a readability score [13]. Intuitively, a higher readability score for a post's title indicates that the title is easier to read, and vice versa. Figure 2 shows the probability of user interactions as the reading ease increases, i.e., as the titles of posts become easier to read because they use shorter words, and smaller sentences. Unsurprisingly, we find that probabilities increase as the readability increases for all types of interactions.

These descriptive statistics provide a first look at how our sample of users browse and vote on Reddit posts. Our next goal is to use these interactivity statistics to predict if and how a user may interact with a new post.

### III. PREDICTING FINE-GRAINED USER ACTIVITY

The user interaction dataset, described in the previous section, contains the positive interactions with Reddit. Development of a fine-grained model of user interactions, however, requires negative training examples, i.e., interactions that the user could have, but did not make. Unfortunately, it is difficult to exactly capture the posts that a user read, but did not click. Instead, we approximate a user's non-interactions by finding posts that were submitted to subreddits that a user often engages with during times when the user was active on Reddit, but that the user did not engage with. Specifically, for each interaction by user $x$ on a post in a subreddit, we found all other interactions in our dataset that occurred 12 hours before or after on posts within the same subreddit from all users except user $x$. Then we removed any posts user $x$ had interactions with. The result is an approximation of the posts that a user likely saw but chose not to interact with. Across all users we collected 325,764 non-interactions, and added them to the user model.

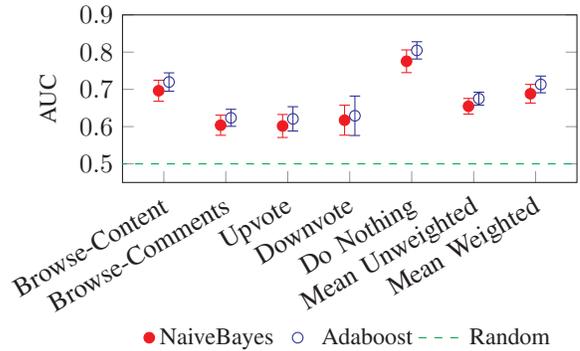

Fig. 3. Mean AUC results with 95% confidence intervals for predictive models using all features of interest, for each interaction type, or a mean of means across all interactions types with each interaction type weighted equally (denoted *mean unweighted*) or proportionally by the user's distribution of interaction types (denoted *mean weighted*).

This culminates in a multi-class classification problem with outputs of browse-content, browse-comments, upvote, downvote, or do-nothing for each post that we estimate a user has seen. We model each user separately and use the first 50% of user-interactions to train a classifier; the remaining 50% are held out for testing.

The feature space includes the post's rank, score, and readability score described in the previous section. We also include the number of subscribers to the post's subreddit, the length of the post's title, and semantic interactivity scores based on the title of the post. The semantic interactivity scores use word vectors to model the ideas and topics that are most correlated with a specific type of interaction. To calculate this score we assign a word vector to each word in the post's title from GloVe, which contains 300d-word vectors pre-trained on 6 billion tokens from Wikipedia and the 5th edition of Gigaword [14]. Scores were trained for each user and post using a multilayer perception classifier that uses ADAM to solve a log-loss function. The loss function considers, for each user, post, and interaction-type, the average word embeddings of the post's title, the average word embeddings of all words in all titles in the dataset, the average word embeddings of titles of posts with the interaction-type, and the average word embeddings of titles of posts that do not have the interaction-type. Simply put, the semantic interactivity scores model a user's probability to interact (or not-interact) with a post based solely on the words used in the title.

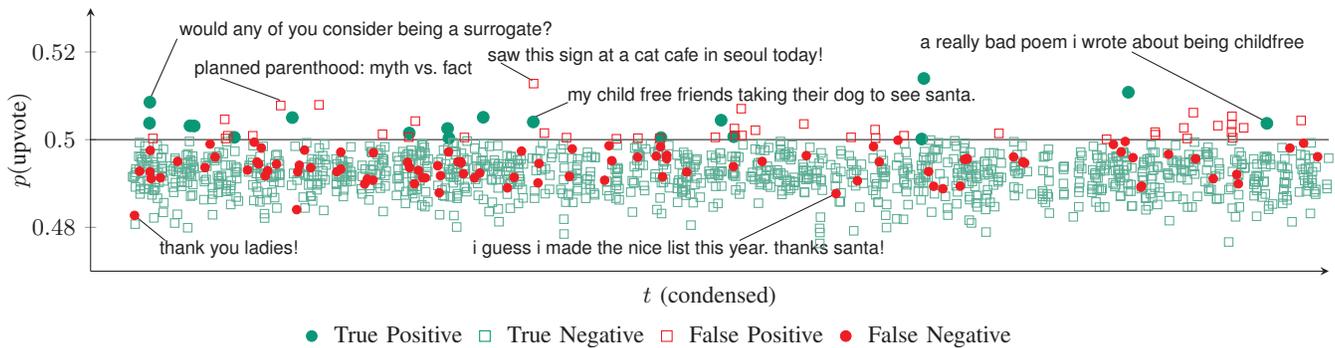

Fig. 4. Upvote predictions for an example user. Y-axis is the predicted class probabilities from the ensemble of classifiers in Adaboost where only interactions above 0.5 are classified as true. Time is to scale, but condensed to fit.

Each user-model is trained using a Naive Bayes or Adaboost classifier. The Adaboost classifier used 100 estimators. Each model for each user was tested against the held-out data for each user. Most users do not interact with most posts, and even when a user does interact with a post they don't interact in all possible ways. We therefore report the area under the ROC (AUC) for each individual interaction-type in Fig. 3. Despite the difficulty of predicting the specific interaction of a specific user on a specific Reddit-post, we find that our models perform reasonably well for all interactions types.

## IV. Discussion

In the present work, we have illustrated the presence of ranking, score, and recency biases in user-interactions with posts. We also find a correlation between the tendency to interact with a post and the post's readability. Surprisingly, we see a smaller ranking bias effect for interactions on the frontpage or /r/all than for interactions on specific subreddits. The difference in ranking bias effect may be because the frontpage is more likely to be filled with high-quality or interesting content than specific subreddits.

We were able to extract several basic ranking- and semantic-post features from the user interactions. The user-models relied on these features to predict fine-grained user-interactions with surprising accuracy. Figure 4 illustrates the held-out test set of an example user where specific true-positives, true-negatives, false-positives and false-negatives are indicated. This example user had an upvote-AUC of .83. Precision and recall for this user are 15% and 29% respectively, whereas a random classifier results in a precision of 0.6% and recall of 3.8%.

We also performed feature ablation tests to determine the usefulness of $2^6$ different feature combinations. Although not illustrated here, we find that, as expected, the inclusion of most (or all) features results in highest accuracy; models containing only 1 or 2 features performed poorly. Top performing combinations contained post-rank, semantic interactivity scores, post-score and the subreddit subscription count. The post title length and reading ease score were found to be the least informative features.

Ultimately, these results show that the history of user-interactions on a social media Web site contains enough information to accurately predict a user's future interactions.